\documentclass[11pt,a4paper]{article}                           

\usepackage{a4}          % Remove before submission

\usepackage[UKenglish]{babel}
\usepackage{amsmath}
\usepackage{amssymb}
\usepackage[usenames]{color} 
\usepackage{multirow}
\usepackage{graphicx}
\usepackage{epstopdf}
\usepackage{array}
\usepackage{hhline}
\usepackage{microtype}
\usepackage[bf]{caption}
\usepackage{color}
\usepackage[usenames,dvipsnames]{xcolor}
 
\usepackage{ifpdf}
\ifpdf
  \usepackage[pdftex,
    pdftitle={SU(5) Tops with (Multiple) U(1)'s},
    pdfauthor={},
    pdfsubject={F-theory compactifications},
    bookmarksopen, bookmarksnumbered, bookmarksopenlevel=2]{hyperref}
\fi
\def\hybrid{\topmargin -20pt    \oddsidemargin 0pt
        \headheight 0pt \headsep 0pt
        \textwidth 6.25in       % A4 paper
        \textheight 9 in       % A4 paper
        \marginparwidth .875in
        \parskip 5pt plus 1pt 
          \jot = 1.5ex
   }
\hybrid
\numberwithin{equation}{section}
\numberwithin{table}{section}

\newcommand{\beq}{\begin{equation}}  \newcommand{\eeq}{\end{equation}}
\newcommand{\bal}{\begin{aligned}}   \newcommand{\eal}{\end{aligned}}
\newcommand{\bea}{\begin{eqnarray}}  \newcommand{\eea}{\end{eqnarray}}

\usepackage{cancel}

\newcommand{\be}{\begin{equation}}
\newcommand{\ee}{\end{equation}}

\begin{document}

\baselineskip=14pt
\parskip 5pt plus 1pt

\vspace*{-1.5cm}
\begin{flushright}    % Publication numbers
  {\small
  }
\end{flushright}

\vspace{2cm}
\begin{center}        % Main title
  {\LARGE On Natural Inflation and Moduli Stabilisation in String Theory}
\end{center}

\vspace{0.75cm}
\begin{center}        % Authors
Eran Palti
\end{center}

\vspace{0.15cm}
\begin{center}        % Institutes
  \emph{Institut f\"ur Theoretische Physik, Ruprecht-Karls-Universit\"at, \\
             %Philosophenweg 19, 69120 
             Heidelberg, Germany}
             \\[0.15cm]
 
\end{center}

\vspace{2cm}

%%%%%%%%%%%%%%%%%%%%%%%%%%%%%%%%%%%%%%%%%%%%%%%
%%%%%%%%%%%%%%%%%%%%%%%%%%%%%%%%%%%%%%%%%%%%%%%
%%%%%%%%%%%%%%%%%%%%%%%%%%%%%%%%%%%%%%%%%%%%%%%
%%%%%%%%%%%%%%%%%%%%%%%%%%%%%%%%%%%%%%%%%%%%%%%
%%%%%%%%%%%%%%%%%%%%%%%%%%%%%%%%%%%%%%%%%%%%%%%
%%%%%%%%%%%%%%%%%%%%%%%%%%%%%%%%%%%%%%%%%%%%%%%
%%%%%%%%%%%%%%%%%%%%%%%%%%%%%%%%%%%%%%%%%%%%%%%
%%%%%%%%%%%%%%%%%%%%%%%%%%%%%%%%%%%%%%%%%%%%%%%

\begin{abstract}

Natural inflation relies on the existence of an axion decay constant which is super-Planckian. In string theory only sub-Planckian axion decay constants have been found in any controlled regime. However in field theory it is possible to generate an enhanced super-Planckian decay constant by an appropriate aligned mixing between axions with individual sub-Planckian decay constants. We study the possibility of such a mechanism in string theory. In particular we construct a new realisation of an alignment scenario in type IIA string theory compactifications on a Calabi-Yau where the alignment is induced through fluxes. Within field theory the original decay constants are taken to be independent of the parameters which induce the alignment. In string theory however they are moduli dependent quantities and so interact gravitationally with the physics responsible for the mixing. We show that this gravitational effect of the fluxes on the moduli can precisely cancel any enhancement of the effective decay constant. This censorship of an effective super-Planckian decay constant depends on detailed properties of Calabi-Yau moduli spaces and occurs for all the examples and classes that we study. We expand these results to a general superpotential assuming only that the axion superpartners are fixed supersymmetrically and are able to show for a large class of Calabi-Yau manifolds, but not all, that the cancellation effect occurs and is independent of the superpotential. We also study simple models where the moduli are fixed non-supersymmetrically and find that similar cancellation behaviour can emerge. Finally we make some comments on a possible generalisation to axion monodromy inflation models.
\end{abstract}

\thispagestyle{empty}
\clearpage

\setcounter{page}{1}

%%%%%%%%%%%%%%%%%%%%%%%%%%%%%%%%%%%%%%%%%%%%%%%
%%%%%%%%%%%%%%%%%%%%%%%%%%%%%%%%%%%%%%%%%%%%%%%
%%%%%%%%%%%                 %%%%%%%%%%%%%%%%%%%
%%%%%%%%%%%  DOCUMENT BODY  %%%%%%%%%%%%%%%%%%%
%%%%%%%%%%%                 %%%%%%%%%%%%%%%%%%%
%%%%%%%%%%%%%%%%%%%%%%%%%%%%%%%%%%%%%%%%%%%%%%%
%%%%%%%%%%%%%%%%%%%%%%%%%%%%%%%%%%%%%%%%%%%%%%%
%%%%%%%%%%%%%%%%%%%%%%%%%%%%%%%%%%%%%%%%%%%%%%%

\newpage

\tableofcontents

%%%%%%%%%%%%%%%%%%%%%%%%%%%%%%%%%%%%%%%%%%%%%%%%%%%%%%%%%%%%%%%%%%%%%%%%%%%%%
\section{Introduction}
%%%%%%%%%%%%%%%%%%%%%%%%%%%%%%%%%%%%%%%%%%%%%%%%%%%%%%%%%%%%%%%%%%%%%%%%%%%%%

Large field inflation is an interesting idea from both experimental and theoretical perspectives. On the experimental side the search for primordial tensor modes is an active field with the feasable target of either detecting tensor modes or placing strong constraints on them in the immediate future. Current results already place strong constraints on large field inflation models implying a close interaction between theory and experiment \cite{Ade:2015lrj}. On the theoretical side large field inflation sits on the boundary between effective field theory and quantum gravity physics. In particular its intrinsic property of super-Planckian field displacements requires a good understanding of quantum gravity corrections. This is therefore an ideal topic to study within string theory since it is physics, with experimental ties, which demands to be framed within the context of a well understood ultraviolet theory. 

Perhaps the most natural way to control the inflaton potential over super-Planckian field ranges is by using a symmetry. In natural inflation we consider the inflaton to be an axion with a continuous shift symmetry that is broken by the inflaton potential to a discrete one \cite{Freese:1990rb}. Natural inflation models in themselves are strongly constrained by current experimental data \cite{Ade:2015lrj}.\footnote{However it is possible that accounting for heavy background fields the induced potential becomes flattened allowing for a better fit to data. See for example \cite{Dong:2010in,Ibanez:2014swa,Buchmuller:2015oma,Dudas:2015lga,Kappl:2015pxa} for an incomplete list of varying realisations of this mechanism. There are also further possibilities not involving flattening \cite{Peloso:2015dsa}.} Independently of how natural inflation models fit the current data they are ideal testing grounds for the question of the possibility of large field inflation as a mechanism within a quantum theory of gravity. They are the class of large field inflation models for which one is able to make the sharpest theoretical statements, for example regarding a possible embedding in string theory. In this sense they are always worth exploring from a theoretical perspective as indicators of whether large field inflation is possible in nature at all. This motivation forms the primary one for this work.

Natural inflation at first appears an attractive proposition in terms of embedding in string theory because axions arise naturally. However a search over string vacua reveals that it does not appear possible for the decay constants of string theory axions to be super-Planckian within a controlled regime \cite{Banks:2003sx,Svrcek:2006yi}, and so the required field range for large field inflation is not available.\footnote{This survey is of course not a proof. One could still attempt to find individual super-Planckian decay constants in string theory. See for example \cite{Kenton:2014gma} for recent work in this direction.} A possible way to circumnavigate this obstruction was suggested in \cite{Kim:2004rp} and is called the Kim-Nilles-Peloso (KNP) mechanism. The basic idea is to consider a two dimensional axion field space which means that, even if each axion has a sub-Planckian decay constant, there is an arbitrarily long field path available. The idea is then to fix one combination of axions with some mechanism at a higher scale and leave the effective light combination as one which is only periodic after a field path much longer than each of the individual decay constants. Or in other words the effective decay constant for the light axion combination is enhanced with respect to the original ones and can be super-Planckian. 

In the low energy effective theory we have a super-Planckian axion decay constant and so we might suspect that this theory is inconsistent from a quantum gravity perspective. This can be related quite nicely to a criterion on effective field theories that can be embedded in quantum gravity called the Weak Gravity Conjecture (WGC) \cite{ArkaniHamed:2006dz}. The WGC is a statement about particle masses and charges in an effective field theory, but this can be mapped to a statement on axions and instanton actions \cite{ArkaniHamed:2006dz,Cheung:2014vva,Rudelius:2015xta,Brown:2015iha,Montero:2015ofa}. The weak form of the WGC states that for any axion there must always be an instanton where the axion appears with a decay constant that is sub-Planckian, let us call this the sub-Planckian instanton. However it does not state what the action of this instanton has to be, and so in principle the action can be very large such that this effect becomes an insignificant modification of the low-energy potential \cite{Rudelius:2015xta,Brown:2015iha,Hebecker:2015rya,Brown:2015lia,Heidenreich:2015wga,Junghans:2015hba}. The strong form of the WGC states further that this sub-Planckian instanton must have an action less than that of the super-Planckian instanton responsible for the inflaton potential, and so it forms the dominant contribution to the potential. It is not clear if the strong, weak, or any version of the WGC holds in nature or string theory. If they do though then the weak version does not strongly constrain natural inflation models while the strong version rules them out. So far it appears that in string theory the strong form is always respected, but in the KNP mixing scenario there is no immediately apparent string theory mechanism that seems to forbid a very large sub-Planckian instanton action.

The purpose of this work is to study the question of the size of the instanton actions and axion decay constants in the context of a well controlled string theory construction. Rather than constructing a full model of inflation, something which typically requires sufficiently complicated constructions that these questions cease to be precisely answerable, we primarily aim for vacua which exhibit the enhancement of effective axion decay constants. We are particularly interested in fully accounting for the physics which is inducing the alignment between the axions. In the original KNP proposal the alignment was induced through a periodic potential for the axions, a second instanton. However in such a situation it is not completely clear how to account for the large parameter which induces the mixing in the full setup. In \cite{Hebecker:2015rya} a string theory variant of the KNP mechanism was proposed where the axion alignment is produced through background string fluxes. This is more promising in terms of keeping track of the alignment parameter since the effect of fluxes on the background has been well studied in the context of moduli stabilisation. Note that the flux induced mixing decompactifies the heavy axionic direction, and so is qualitatively different from the KNP setup in this respect, but the underlying result of an effective enhanced axion decay constant for the light direction remains the same. The proposal of \cite{Hebecker:2015rya} is in a type IIB string theory setting which is very rich and promising phenomenologically, but on the other hand is rather difficult to analyse quantitatively. In particular the effect of the fluxes on the background is not straightforwardly manifest. In this paper we will present a new realisation of the flux mixing mechanism which is significantly simpler to analyse and where the physics of the flux background will be more transparent. 

The main type of vacua we will consider are compactifications of type IIA string theory on Calabi-Yau manifolds with background fluxes. The moduli stabilisation aspects of these vacua were first studied in \cite{DeWolfe:2005uu}. The particular property which we are interested in is that it is possible to fix all the moduli in a sufficiently simple way that the moduli values can be solved for in terms of the fluxes. Further, although all the geometric moduli are fixed, axionic fields remain which are then lifted by non-perturbative effects. Since the geometric moduli are fixed, the magnitude of these non-perturbative effects can be precisely solved for in terms of the fluxes. Finally the axionic fields which remain perturbatively massless are linear combinations of the original `fundamental' axions, where the light linear combination is determined in terms of the flux background. We therefore have all the ingredients of an axion alignment mechanism within a well understood string theory setting. The crucial property of these vacua is that they allow us to make on-shell statements regarding the magnitude of the decay constants and instantons. The reason this is a key property is that decay constants and instantons, like any coupling in string theory, are moduli dependent quantities and so to study them in an ultraviolet context we must account for the effect of the fluxes on their values. As stated, these are not candidate inflation vacua in themselves as yet, in particular since for simplicity we will primarily consider supersymmetric vacua which are four-dimensional AdS space. But they will provide us with insights into the questions regarding the interplay between axion decay constants and quantum gravity effects, which in turn have crucial implications for models of large field inflation.

The paper is formed as follows. In section \ref{sec:cyflux} we review and extend the existing work on perturbative moduli stabilisation in type IIA string theory on a Calabi-Yau with fluxes. In section \ref{sec:npeff} we add the non-perturbative effects which fix the remaining axions. In section \ref{sec:fluxdec} we then combine all the data to study the possibility of inducing super-Planckian effective axion decay constants and study the magnitude of the instanton corrections to the resulting axion potential. In section \ref{sec:genanal} we generalise our results to arbitrary superpotentials. In section \ref{sec:nonsusy} we consider a simplified setup in which we can study non-supersymmetirc vacua. We summarise our findings in section \ref{sec:summary}.  In appendix \ref{sec:axionmon} we present a toy model which captures some back-reaction effects of the fluxes in the case of an axion monodromy model.

%%%%%%%%%%%%%%%%%%%%%%%%%%%%%%%%%%%%%%%%%%%%%%%%%%%%%%%%%%%%%%%%%%%%%%%%%%%%%
\section{Type IIA moduli stabilisation on a Calabi-Yau with flux}
\label{sec:cyflux}
%%%%%%%%%%%%%%%%%%%%%%%%%%%%%%%%%%%%%%%%%%%%%%%%%%%%%%%%%%%%%%%%%%%%%%%%%%%%%

In this section we consider moduli stabilisation in the context of a CY with flux setup. We begin be reviewing some of the main results presented in \cite{DeWolfe:2005uu} and then move on to a more explicit stabilisation scenario on a particular CY manifold. 

We begin by summarising the effective four-dimensional supergravity theory following \cite{Grimm:2004ua} and using the conventions in \cite{Palti:2008mg}. 
The Kahler moduli sector arises from expanding $J_c = B+iJ$ in a basis of harmonic orientifold odd $(1,1)$ forms $\omega_i$ which leads to the scalar components of the chiral superfields labeled by $T_i = b_i+it_i$. The complex structure moduli arise from expanding the holomorphic three-form and RR three-form combination $\Omega_c = C_3 + 2 i \mathrm{Re}\left(C \Omega\right)$. The real compensator field $C$ incorporates the dependence on the ten-dimensional dilaton $\phi$ through $C =\frac{1}{\sqrt{8}}e^{-\phi}e^{-K^K/2+K^{cs}/2}$ where the Kahler potentials for the Kahler and complex structure moduli are defined below. Here the relevant basis of harmonic three-forms are orientifold even ones $\left\{\alpha_k,\beta_{\lambda}\right\}$ and orientifold odd ones $\left\{\alpha_{\lambda},\beta_k\right\}$, where the indices $\lambda$ and $k$ range over the appropriate number of forms. It is always possible to choose, through an appropriate symplectic transformation, an index range where $k=0$. We will therefore have the basis of forms
\be
\left\{\alpha_0,\beta^{\lambda}\right\} \in H^3_+ \;,\;\;\; \left\{\alpha_{\lambda},\beta^0\right\} \in H^3_- \;.
\ee
We can expand the holomorphic three-form in terms of its period structure, which after imposing the orientifold projection \cite{Grimm:2004ua}, takes the form
\be
C\Omega =\mathrm{Re} \left(C Z^0\right) \alpha_0  + i \mathrm{Im} \left(C Z^{\lambda}\right) \alpha_{\lambda} 
-\mathrm{Re} \left(C F_{\lambda}\right) \beta^{\lambda}  - i \mathrm{Im} \left(C F_0\right) \beta^0 \;.
\ee
We further expand the RR three-form as
\be
C_3 = - \sigma \alpha_0 - \nu_{\lambda} \beta^{\lambda} \;.
\ee
The four-dimensional superfields arise from the expansion
\be
\Omega_c = i S \alpha_0 + i U_{\lambda} \beta^{\lambda} \;,
\ee
with components $S = s + i \sigma$ and $U_{\lambda} = u_{\lambda} + i \nu_{\lambda}$, where 
\be
s = 2\mathrm{Re}\left(C Z^0\right) \;, \;\; u_{\lambda} = - 2 \mathrm{Re}\left(C F_{\lambda}\right) \;.
\ee
The Kahler potential for the resulting four-dimensional theory is given by 
\be
K = - \log 8 {\cal V} 
- 2\log{2\left[\mathrm{Im} \left(CZ^{\lambda}\right) \mathrm{Re} \left(C F_{\lambda} \right) - \mathrm{Re} \left(CZ^0\right) \mathrm{Im} \left(CF_0\right) \right]}
\ee
In the large complex structure limit the prepotential defining the periods takes the form
\be
F = \frac16 d_{\lambda \rho \sigma} \frac{Z^{\lambda} Z^{\rho} Z^{\sigma}}{Z^0} \;,
\ee
which leads to the Kahler potential \cite{Grimm:2004ua,Palti:2008mg}
\be
K = - \log 8 {\cal V} - \log \left(S+\overline{S}\right) - 2 \log {\cal V}' \;, \label{kahleriia}
\ee
where we have
\bea
{\cal V} &=& \frac16 k_{ijk} t^i t^j t^k \;,\;\; {\cal V}' = \frac16 d_{\lambda \rho \sigma} v^{\lambda} v^{\rho} v^{\sigma} \;, \\ \nonumber
 v^{\lambda} &=& - 2 s^{\frac12} \mathrm{Im} \left(C Z^{\lambda} \right) \;, \;\; u_{\lambda} = \partial_{v^{\lambda}} {\cal V}' \;,
\eea
with $k_{ijk}$ being the triple intersection numbers of the CY. The Kahler potential (\ref{kahleriia}) will be the final expression we will work with. The parameterisation of the complex-structure moduli differs from that in \cite{DeWolfe:2005uu} but it is the more suitable one for explicitly solving for the moduli. It also makes the mirror symmetry to type IIB with O3/O7 planes manifest: the moduli $v^{\lambda}$ are the duals of the 2-cycle volumes in IIB while the $u_{\lambda}$ are mirror to the 4-cycle volumes forming the superfields.

A classical superpotential is generated in the presence of fluxes. These are expanded as
\bea
H_3 &=& q^\lambda \alpha_\lambda - h_0 \beta^0 \,, \quad F_0 = -f_0 \,, \quad F_2 = - \tilde{f}^i \omega_i \nonumber \\
 \quad F_4 &=&- f_i \tilde{\omega}^i \,, \;,\quad F_6 =- \tilde{f}_0 \epsilon \;,
\eea
where $\tilde{w}^i$ are the four-form duals of the $w_i$ and $\epsilon$ is the CY volume form.
The resulting perturbative superpotential, in the large volume limit, reads \footnote{Away from the large volume limit the superpotential receives also $\alpha$' corrections as studied in \cite{Palti:2008mg}. Note however that there are no corrections to the superpotential coming form moving away from the large complex-structure limit. Rather these manifest themselves in the definitions of the $u_{\lambda}$ in terms of the $v^{\lambda}$.}
\be
W = \frac{f_0}{6} k_{ijk} T^i T^j T^k + \frac12 k_{ijk} \tilde{f}^i T^j T^k - f_i T^i + \tilde{f}_0 - i h_0 S - i q^{\lambda} U_{\lambda} \;. \label{suppot}
\ee
The solution of the F-terms for this system was studied in \cite{DeWolfe:2005uu} where it was shown that generically all the complex-structure and Kahler moduli are fixed, also the dilaton and the axion partners of the Kahler moduli, and finally one single axionic direction from the set of axions $\left\{\sigma,\nu_{\lambda}\right\}$. The resulting supersymmeric minimum was AdS. We will repeat this analysis and extend it by providing further detail on the precise values of the complex-structure moduli. 

Consider first the F-term for the axio-dilaton $S$. This gives us the two equations
\be
h_0 \sigma + q^{\lambda} \nu_{\lambda} = - \mathrm{Re} W^T \;,\;\;\; 2 h_0 s = - \mathrm{Im} W \;,
\ee
where $W^T$ refers to the part of the superpotential which does not depend on $S$ and $U_{\lambda}$.
The first of these fixes the single combination of RR axions, while the second can be taken to fix the dilaton. Now consider the $U_{\lambda}$ F-terms, these give rise to only $\lambda$ real equations which can be taken to fix the complex-structure moduli
\be
\frac{q^{\lambda}}{K_{u_{\lambda}}} = \frac12 \mathrm{Im} W = - h_0 s\;. \label{UFterms}
\ee
Our primary interest will be in equations (\ref{UFterms}), but first let us discuss the Kahler moduli sector. 

It is possible to show that the equations (\ref{UFterms}), combined with the fact that ${\cal V}'$ is homogeneous of degree $\frac32$ in the $u_{\lambda}$, imply that the Kahler moduli F-terms take the form
\be
\partial_{T^i} W^T - i \left(\partial_{T^i}K \right) \mathrm{Im} W^T = 0 \;.
\ee
This means that they are independent of the values of the complex-structure moduli or the dilaton, as well as the NS flux. The stabilisation of the Kahler moduli proceeds identically to the analysis of \cite{DeWolfe:2005uu} and is not of much interest for our purposes. It suffices to state that the moduli are fixed as solutions to a quadratic equation involving the fluxes. It is worth noting though that the value of the superpotential in the vacuum is fixed by this procedure to be
\be
\mathrm{Im} W = -\frac{12}{15} f_0 {\cal V} \;.
\ee
We will assume that for a sufficient choice of fluxes the Kahler moduli can be fixed at large enough values to control the $\alpha'$ expansion, and treat the volume ${\cal V}$ as a free parameter. The resulting absolute value of the dilaton is fixed as
\be
s = \frac{2 f_0}{5 h_0} {\cal V} \;. \label{dilatonvev}
\ee

In order to understand how the complex-structure moduli are fixed in more detail we have to restrict to a particular CY. The simplest example is a single modulus one, for example the mirror of the quintic. In this case we have
\be
{\cal V}'  = u_1^{3/2} \;
\ee
which implies that the moduli are stabilised in the ratio
\be
\frac{s}{u_1} = \frac{q^1}{3h_0} \;.
\ee
A less trivial example is the mirror CY to the $P_{[1,1,1,6,9]}$ manifold, as studied for example in \cite{Denef:2004dm}. For this particular example we have 2 complex-structure moduli, and the complex-structure volume takes the form
\be
{\cal V}'  = u_1^{3/2} - u_2^{3/2} \;.
\ee
For this case we find the following ratios
\be
\frac{s}{u_1} = \frac{\left(q^1\right)^3 + \left(q^2\right)^3}{3 h_0 \left(q^1\right)^2} \;,\;\;\; \frac{s}{u_2 } = \frac{\left(q^1\right)^3 + \left(q^2\right)^3}{3 h_0 \left(q^2\right)^2} \;,\;\;\;\;
\sqrt{\frac{u_1}{u_2}} = -\frac{q^1}{q^2}\;. \label{suratio}
\ee

The last bit of information we need before proceeding to analyse the interaction with non-perturbative effects is the constraint on the fluxes coming from tadpoles. These take the form \cite{DeWolfe:2005uu}
\be
f_0 h_0 = -Q^0_{D6} \;,\;\;\; f_0 q^{\lambda} = -Q^{\lambda}_{D6} \;. \label{tadpoles}
\ee
Here $Q_{D6}$ stands for the induced D6 charge from local sources, O6-planes and D6-branes, wrapped on the 3-cycle orthogonal to the 3-cycle that the NS flux is threading (so as to act as sources/sinks for it). The precise values of the local contribution depends on the particular orientifold projection imposed, which fixes the O6-planes, and the number of D6 or anti-D6 branes added on top. Note that an orientifold plane can contribute at most ${\cal O}\left(100\right)$ negative D6 charge to the tadpoles on any of the cycles. The precise number can vary depending on the manifold and projection, but it is never parametrically large. 
The sign of the fluxes is a priory arbitrary but gets restricted upon moduli stabilisation. In particular from (\ref{dilatonvev}) we see that the sign of $f_0 h_0$ must be positive in a physical vacuum and therefore the fluxes contribute positive D6 charge to the background. Similarly, from (\ref{suratio}) we learn that $q^1$ and $q^2$ must have opposite sign, that $\left|q^1\right| > \left|q^2\right|$, and that $f_0 q^1$ must be positive. Therefore the fluxes contribute positive D6 charge on the big cycle $u_1$ and negative D6 charge on the small cycle $u_2$. 

It is interesting to note that $q^1$ can not be parametrically large while maintaining supersymmetry of the background. After saturating any orientifold planes charge increasing $q^1$ must be accompanied by introducing anti-D6 branes. As we will show in the next section $q^1$ is precisely the parameter that controls the enhancement of the axion decay constant in the implementation of the KNP mechanism. Could the tadpole constraints therefore be placing a fundamental string theory limit on the size of $q^1$ and thereby censoring the axion decay constant enhancement? Possibly yes, but probably no. First note that adding anti-D6 branes is still consistent in string theory and one does not expect that a mechanism censoring super-Planckian axion decay constants in string theory should be tied to the supersymmetry of the background. Secondly, $q^1$ can still be of ${\cal O}\left(100\right)$ in appropriate backgrounds while preserving supersymmetry, and so would still provide an axion decay constant enhancement. While as we will argue in the following sections, for this model the moduli space properties imply that there can be no enhancement at all of the effective axion decay constant beyond the size of one of the original ones, ie. the effect of the flux drops out completely. This therefore appears to be a stronger constraint than any constraint coming from the tadpoles. 

Finally, it is possible to consider other string backgrounds where additional metric fluxes can balance the flux contribution from the RR fluxes to the tadpoles parametrically \cite{Camara:2005dc}. Indeed it was argued in \cite{Acharya:2006ne} that localised sources deform the manifold to an $SU(3)$-structure one which supports metric fluxes (torsion). Examples of such backgrounds are twisted-torus compactification, as studied for example in \cite{Camara:2005dc}. These are analogous in some ways to the current set-up but differ in three aspects: the four-dimensional theory can be uplifted to a 10-dimensional solution even with localised sources, some of the axions are lost, the flux can be parametrically increased without having to introduce localised source contributions to cancel it. The fact that axions are lost is interesting, and can be understood either as a topology change in the manifold thereby losing cycles, or in terms of the superpotential where one finds a new contribution of the type $W \supset b_{i}^{\lambda}T^i U_{\lambda}$. Therefore for general values of the $b_{i}^{\lambda}$ the number of axions is reduced by the index range of $i$. However by choosing the coefficients $b_{i}^{\lambda}$ proportional to the $q^{\lambda}$ one can reinstate the full set of axions, ie. lead to a superpotential which only depends on one combination of $U_{\lambda}$ independently of the values of the $T^i$.\footnote{If we consider the set-up of \cite{Camara:2005dc}, we have that, in their notation, choosing $b_{ji}=b_i$, $b_{ii}=-b_i$, $a_i=a$, $3h_i a = -h_0 b_i$, as well as some constraints on the RR fluxes (eg. $m=c_2$, $h_0=3a$) solves the tadpoles. In particular the single axion combination which is fixed by these fluxes is $3 a \sigma + \sum_i b_i \nu_i$ which means that we can take the $b_i$ large unconstrained by tadpoles.} Geometrically this means that only one 3-cycle combination develops a boundary, the one dual to $H$, which allows $H$ to cancel against $dF_2$ in the tadpoles exactly thereby requiring no local sources on each individual cycle. It is possible to also consider splitting the fluxes into proportional groups thereby fixing more and more axions up to the maximum when none of the fluxes are proportional. 

%%%%%%%%%%%%%%%%%%%%%%%%%%%%%%%%%%%%%%%%%%%%%%%%%%%%%%%%%%%%%%%%%%%%%%%%%%%%%
\subsection{Non-perturbative effects and application to inflation}  
\label{sec:npeff}
%%%%%%%%%%%%%%%%%%%%%%%%%%%%%%%%%%%%%%%%

The vacuum described in the previous section fixes all the moduli apart from the RR axions. However the superpotential receives further non-perturbative corrections coming from Euclidean D2-brane instantons on orientifold even three-cycles. Note that this is the set of cycles orthogonal to those which support the NS H-flux and so the full configuration is free from Freed-Witten anomalies and is consistent. The instantons fix the remaining axions as long as their number is as many as there are unfixed axions. Note that unlike the more familiar type IIB setting, the geometric part of the complex-structure moduli is fixed already perturbatively by the fluxes, and so the magnitude of the instanton action is determined in terms of flux ratios. This allows for good control over the magnitude of the instantons and ensures that an exponential hierarchy in mass can be induced between the remaining axion stabilisation and the rest of the moduli. This is a useful feature for potential inflation models where the axion inflaton can be made lighter than the rest of the moduli. It also implies that we can treat the axion stabilisation after integrating out the more massive moduli. The non-perturbative superpotential takes the form 
\be
W^{NP} = W^P + \sum_{I} A_{I} e^{- a^I_0 S - a^I_{\lambda} U_{\lambda}} \;,
\ee
where $W^P$ is the perturbative superpotential. Here the index $I$ runs over the number of different instantons, while the constants $a^I_{0}$ and $a^I_{\lambda}$ refer to the combination of 3-cycles wrapped by the instanton.\footnote{For simplicity we will henceforth consider only the case of instantons as opposed to gaugino-condensation on D-brane stacks.} Since there is a potentially exponential hierarchy between the axions masses and the moduli masses we can integrate out the moduli and work with the effective theory for the axions. The resulting effective theory is of the type considered when studying models of natural inflation with multiple axions. 
Let us consider the example studied in the previous section with two complex-structure moduli. The effective potential takes the schematic form
\be
V_{\mathrm{eff}} = V_0 + A'_0 e^{-s} \left(1-\cos{\sigma}\right) + A'_1 e^{-u_1} \left(1-\cos{\nu_1}\right) 
+ A'_2 e^{-u_2} \left(1-\cos{\nu_2}\right) \;.
\ee
Note that we have included a pure dilaton instanton in the scalar potential. We do not expect that pure dilaton instantons have the correct zero mode structure to contribute to the superpotential because they are mirror to type IIB D(-1) instantons which have too many zero modes. However on the type IIA side the additional zero modes are not obviously manifest, and in the presence of flux the IIB mirror is not simple, and so it may be that they could contribute to the superpotential. In any case such instantons do contribute to the effective theory and in particular upon supersymmetry breaking one expects the additional vector like zero modes to be lifted and the resulting instanton to contribute to the potential \cite{Blumenhagen:2009qh,Montero:2015ofa}.\footnote{Note that also for instantons involving the $u_i$ there is some constraint on when they contribute to the potential. From the tadpoles we should have O6 planes wrapping the even 3-cycles in the same class as the cycles wrapped by the instantons involving the $u_i$. There is then a constraint on when these instantons can contribute to the superpotential amounting to a self intersection of the wrapped cycle away from the orientifold \cite{Blumenhagen:2009qh}, something which generically is expected to occur. In any case, in a cosmological setting with broken supersymmetry they are expected to contribute to the potential anyway.} This is the relevant situation for inflation, and so we will keep track of the presence of these instantons by including them in the expressions for the scalar potential. 

We would now like to realise the KNP mechanism \cite{Kim:2004rp}, or more precisely a variant of it where we fix the light axion direction through flux choices similar to the mechanism of \cite{Hebecker:2015rya}. Consider the single modulus model where we have the two axions $\sigma$ and $\nu_1$, one combination of which is fixed at a high scale by the superpotential. 
The remaining light field $\psi$, which we canonically normalise, is related to the original axions as
\be
\sigma = -\frac{q^1 \psi}{N} \;,\;\; \nu_1 = \frac{h_0 \psi}{N} \;,\;\; N \equiv \sqrt{h_0^2 K_{U_1\bar{U}_1} + \left(q^1\right)^2 K_{S\bar{S}}} \;.
\ee
The effective potential for this light axion is given by 
\be
V_{\mathrm{eff}}\left(\psi\right) = V_0 + A' e^{-s} \left(1 -\cos{ \frac{q^1\psi}{N}} \right) + B' e^{-u_1} \left(1 -\cos{ \frac{h_0\psi}{N}} \right) \;.
\ee
And therefore we have the effective axion decay constants
\be
f^{s}_{\psi} = \frac{N}{q^1}  \;,\;\; f^{u_1}_{\psi} = \frac{N}{h_0} \;.
\ee
This is a realisation of an alignment mechanism in the sense that we can obtain a parametric enhancement of the effective axion decay constant $f^{u_1}_{\psi}$ relative to an original axion decay constant. For example we can write
\be
f^{u_1}_{\psi} = \frac{q^1 \sqrt{K_{S\bar{S}}}}{h_0} \sqrt{\left(\frac{h_0}{q^1}\right)^2 \frac{K_{U_1\bar{U}_1}}{K_{S\bar{S}}} + 1} 
= \frac{q^1 f_{\sigma}}{h_0} \sqrt{\left(\frac{h_0 f_{\nu_1}}{q^1f_{\sigma}}\right)^2  + 1} \;, \label{fscalconst}
\ee
where $f_{\nu_1}$ and $f_{\sigma}$ are the associated axion decay constants for the canonically normalised axions $\nu_1$ and $\sigma$. If we keep $f_{\nu_1}$ and $f_{\sigma}$ fixed then for large flux $q^1$ we obtain a parametric enhancement of $f^{u_1}_{\psi}$ from $f_{\sigma}$. So while $f_{\sigma}$ must be sub-Planckian the effective axion decay constant $f^{u_1}_{\psi}$ may be parametrically increased to possibly super-Planckian values. We will see however that this is not so, but first let us discuss the magnitude of the instanton corrections.

We are now in a position to explicitly consider the corrections coming to the axion potential over the full range of the effective axion decay constant $f^{u_1}_{\psi}$. As discussed in the introduction the weak gravity conjecture implies the existence of instantons whose action is not parametrically enhanced by $q^1$. And in particular they should lead to corrections to the axion potential of order its natural decay constant, in this case $f_{\sigma}=1/2s$. This is the instanton involving the dilaton $s$. The key question is whether these can be suppressed with respect to the leading effect coming from the instantons involving $u_1$. The ratio of the magnitude of these effects is exponential in the ratio (\ref{suratio}). We see that indeed for large $q^1$ this scales as $s/u_1 \sim q^1$ and so $s$ is parametrically larger than $u_1$ implying a strong suppression of the associated instanton effects. Therefore this result provides an explicit example within a string background where the so called `loophole' of the weak version of the weak gravity conjecture \cite{Rudelius:2015xta,Brown:2015iha,Hebecker:2015rya} can be realised and the axion potential can remain suitably flat over the axion decay range. However we will soon show that there is an apparent fundamental obstruction to utilising this scenario to reach super-Planckian decay constants related to the way that the flux which is used to fix the axion combination also affects the decay constants. 

First though let us consider the less trivial setup where we have two complex-structure moduli. The qualitatively new effect is that we have two remaining light axions after fixing the combination of the superpotential. However since $u_1/u_2 = \left(q^1/q^2\right)^2$ we can induce an exponential hierarchy between the masses of these two axions $\nu_1$ and $\nu_2$. Let us simplify the system therefore by taking $q^1 \gg q^2$ and then integrating out the $\nu_2$ axion. What remains is an effective theory with a single axion $\psi$ which is a linear combination of the original three axions $\sigma$, $\nu_1$, $\nu_2$. The effective theory is under good control because we have an exponential hierarchy between the mass scales of the three axions
\be
M\left(h_0 \sigma+q^1\nu_1+q^2\nu_2 \right) \sim 1 \;, \;\;\; M\left(\nu_2\right) \sim e^{-u_2}\;,\;\;\; M\left( \psi \right) \sim e^{-u_1} \;.
\ee
Even with the mass hierarchy, integrating out $\nu_2$ is is not straightforward due to the kinetic mixing it has with the $\nu_1$ axion. However the off diagonal terms in the kinetic matrix are suppressed by powers of $q^2/q^1$ which means that up to corrections in this ratio we can treat the axions $\nu_1$ and $\nu_2$ as orthogonal. In this approximation integrating out $\nu_2$ reduces to the two axion case discussed above. Therefore the resulting effective theory is the same as the two axion one up to corrections in $q^2/q^1$ which will not change the resulting physics qualitatively. 

%%%%%%%%%%%%%%%%%%%%%%%%%%%%%%%%%%%%%%%%%%%%%%%%%%%%%%%%%%%%%%%%%%%%%%%%%%%%%
\subsection{Flux stabilisation of decay constants}
\label{sec:fluxdec}
%%%%%%%%%%%%%%%%%%%%%%%%%%%%%%%%%%%%%%%%%%%%%%%%%%%%%%%%%%%%%%%%%%%%%%%%%%%%%

The expression (\ref{fscalconst}) shows that if we keep the fundamental axion decay constants $f_{\sigma}$ and $f_{\nu_1}$ fixed we obtain a parametric enhancement of the effective decay constant $f_{\psi}$ with the flux $q^1$. However in string theory the axion decay constants are not constant but rather are moduli dependent. In turn the values of the moduli are dependent on the flux $q^1$ since it induces an energy density in the extra dimensions to which the space adjusts. The set-up we have allows us to make this dependence explicit. In particular we have that for the single modulus case
\be
\frac{f_{\nu_1}}{f_{\sigma}} = \frac{q^1}{\sqrt{3}h_0} \;. \label{decayconstrat}
\ee
While for the two modulus case we have
\be
\left(\frac{f_{\nu_1}}{f_{\sigma}}\right)^2 = \frac{2\left(q^1\right)^3 - \left(q^2\right)^3}{6h^2_0 q^1} \;.
\ee
which reduces to (\ref{decayconstrat}) up to corrections in $q^2/q^1$. Now using (\ref{decayconstrat}) we see that
\be
f^{u_1}_{\psi} = \frac{2q^1 f_{\sigma}}{\sqrt{3}h_0} = 2f_{\nu_1} + {\cal O}\left(\left(\frac{q^2}{q^1}\right)^{3}\right) \;.
\ee
But this shows that the effective axion decay constant, although parametrically enhanced with respect to $f_{\sigma}$ is not enhanced with respect to $f_{\nu_1}$. And since the original `fundamental' decay constant $f_{\nu_1}$ can not be super-Planckian, neither can the effective $f_{\psi}$.

What has happened is that holding the $u_1$ modulus fixed the axion decay constant $f_{\sigma}$ scales as $1/q^1$ precisely canceling any enhancement from the alignment. So although the axion period spans many periods of $f_{\sigma}$ they get smaller and smaller. An equivalent statement is that if we hold instead $f_{\sigma}$ fixed, or the dilaton, then the flux drives $f_{\nu_1}$ very large. To see why this is forbidden we can return to the question of why the original `fundamental' decay constant $f_{\nu_1}$ can not be super-Planckian in the first place. Its form in this case is
\be
f_{\nu_1} =  \frac{\sqrt{3}}{2u_1}  + {\cal O}\left(\left(\frac{q^2}{q^1}\right)^{\frac32}\right) \;.
\ee
Now making $f_{\nu_1}$ large implies making $u_1$ small and this drives us to the limit of moduli space. Physically it corresponds to shrinking a 3-cycle in the geometry, and in this case the energy density of the flux which is used for the axion mixing is driving the cycle to this vanishing locus. Once the 3-cycle becomes of order the string scale the effective theory breaks down. In this particular case it is known that in IIA when a 3-cycle shrinks to zero size an infinite number of instantons wrapping it contribute to the moduli space geometry and thereby smooth it out \cite{Ooguri:1996me}. These instantons are just the multi wrappings of the instanton which was responsible for the potential in the first place whose volume is measured by $u_1$. From the perspective of the WGC we see that indeed particles become light if we try to make $f_{\psi}$ super-Planckian but it is not those associated to the dilaton instanton $s$ but to the instanton in $u_1$ and higher copies of it. The multi-wrapped instantons have periodicities reduced by the wrapping number and so reduce the period size as in the mechanism described in \cite{Banks:2003sx}.

At this point it is worth stating some properties of the vacuum solution in which we have observed this effect. First, since the vacuum is supersymmetric a natural question is how can the geometric moduli be fixed while leaving axions unfixed? The reason is that we are in AdS space and so the geometric moduli are sitting at the Breitenlohner-Freedman bound: they have a tachyonic mass but they are completely stable of course. The most important reason for considering the vacuum described is that it accounts for the interaction between the fluxes and the moduli. The fact that it is supersymmetric or AdS is just a byproduct of the simplest method of finding such on-shell vacua which is by solving the F-terms. However it is important to keep in mind that it is possible that some of the physics that we have observed could be due to the supersymmetric or AdS nature of the vacuum. Some arguments against this possibility and towards a more general validity follow. First the data of the vacuum which we used is primarily the vanishing of the F-terms for the geometric partners of the axions. This alone does not require or imply that the full vacuum must be supersymmetric. Secondly, in section \ref{sec:nonsusy} we will present an analysis of a simple model where we will not impose the vanishing of the F-terms for even the superpartners of the axions. A complete analysis of the non-supersymmetric case is much more challenging than the supersymmetric one, so we will not be able to solve for the moduli fully, but will show that similar structures can emerge also in that case. Finally there is the physically intuitive argument: the primary effect observed here is that the moduli are adjusting to the energy density of the flux which causes the mixing. This interaction does not appear dependent on supersymmetry or the value of the cosmological constant.

%%%%%%%%%%%%%%%%%%%%%%%%%%%%%%%%%%%%%%%%%%%%%%%%%%%%%%%%%%%%%%%%%%%%%%%%%%%%%
\section{General analysis of F-term fixing of moduli}
\label{sec:genanal}
%%%%%%%%%%%%%%%%%%%%%%%%%%%%%%%%%%%%%%%%%%%%%%%%%%%%%%%%%%%%%%%%%%%%%%%%%%%%%

How can we avoid the scenario in the example above? One could contemplate using the $u_2$ modulus for enhancement with large $q^2$, however we must have $u_1>u_2$ and so the axion mass hierarchies are fixed to be as used above. What about more complicated Calabi-Yau manifolds? To see how the mechanism generalises consider an arbitrary superpotential and Kahler potential such that there is an axion flat direction, which means that the Kahler potential is independent of the axion while the superpotential is independent of the full superfield involving the axion. In particular we want to consider a superpotential dependent on only a combination of superfields
\be
K = K\left(u_\lambda\right) \;,\;\; W = W\left(\sum_\lambda q^\lambda U_\lambda \right) \;,\;\;\; U_\lambda = u_\lambda + i \nu_\lambda \;.
\ee
Note that we should differentiate between the requirement of the superpotential to be independent of the axion off shell, as we do, and the possibility that we have an axion in a certain vacuum only. An example of vacua in which such `on-shell' axions appear are found in \cite{Camara:2005dc} for example. However such axions are not expected to have a well protected shift symmetry, especially once supersymmetry is broken during inflation.
The imaginary part of the F-term equations takes the form
\be
\mathrm{Im} W_{U_\lambda} + \frac12 K_{\lambda} \mathrm{Im} W = 0 \;, \label{super1}
\ee
where we define $K_{\lambda}=\partial_{u_\lambda} K$. In particular this means
\be
\frac{K_{\lambda}}{K_{\sigma}}= \frac{q^\lambda}{q^\sigma} \;. \label{kahler1}
\ee
The equation (\ref{kahler1}) is satisfied for any of the axions that are involved in the mixing. In the examples studied in the previous section the Kahler potential for $s$ and $u_1$ had the property that up to small corrections
\be
\left(K_{u_{\lambda}}\right)^2 \sim K_{u_{\lambda}u_{\lambda}} \;. \label{kahler2}
\ee
Combining this Kahler potential behaviour with (\ref{kahler1}) directly implies 
\be
\frac{f_{\lambda}}{f_{\sigma}} \sim \frac{q^{\lambda}}{q^{\sigma}} \;, \label{fratios}
\ee
which implied that the parametric enhancement only occur with respect to one but not both of the `fundamental' axion decay constants, which tells us that the effective decay constant should be sub-Planckian.   

How generic is the property (\ref{kahler2}) of the Kahler potential. For the dilaton it is always satisfied at tree-level. For the geometric moduli it is satisfies precisely in a torodial setting since $K^{\mathrm{Torodial}} \sim \log u_{\lambda}$. In a CY setting however it is certainly not satisfied by all the moduli. The type of moduli which do satisfy it approximately are so called `large' moduli which measure the volumes of the whole CY. Note that henceforth, to be more intuitive, we will refer to the mirror moduli which measure cycles volumes. More generally the condition amounts to the statement that in the vacuum the moduli are fixed in such a way that in a Taylor expansion of the CY volume in the modulus $u_{\lambda}$ is such that the term with the largest power of $u_{\lambda}$ dominates in magnitude over the other terms. This is a generic expected behaviour for `large' moduli, an example of which is the volume cycle $u_1$ above. 

In the setup of type IIA on Calabi-Yau with fluxes the lightest $\nu_{\lambda}$ axion will always be the one associated to the largest valued modulus. Therefore, at least for that modulus, we generically expect behaviour like (\ref{kahler2}), which is as was the case for $u_1$ in the example. Further, unless the dilaton flux parameter $h_0$ is the one being dialed large, we generically expect that the dilaton is larger in magnitude that any $u_{\lambda}$ since it goes like $s \sim \frac{q^{\lambda} {\cal V}'}{h_0\left(\partial_{u_{\lambda}}{\cal V}'\right)}$. We will therefore have an effective theory where we integrate out all the heavier axions fixed by the more dominant instanton effects leading to the $\sigma-\nu_1$ axion situation as described in the main example. Therefore the system studied in the example is quite general within this IIA on Calabi-Yau setting up to the assumptions above. However exceptions can occur and we will present in the next section an example which is not of this class due to $h_0$ being the large flux parameter.

Can we use what we have learned to constrain other type of compactifications than IIA on CY? If we still insist on fixing the axion superpartners with vanishing F-terms then the most significant new freedom we have for arbitrary superpotential is that we may mix any two axions, not necessarily the superpartners of the dilaton and largest modulus, to enhance the effective decay constant of the lightest field. One realisation of this is to consider backreacting the fluxes and deforming the CY in the IIA setup to an SU(3)-structure manifold. Then, as discussed previously, we can fix a number of independent axion combinations depending on the form of the metric and RR flux. Another realisation is the scenario of \cite{Hebecker:2015rya} where, since the superpotential is cubic in the axions, all the axions are fixed generically and some flux tuning is required to keep flat directions. We can also consider a type IIB setup where the dilaton is fixed perturbatively while the Kahler moduli are fixed non-perturbatively (as studied in \cite{Long:2014dta,Rudelius:2015xta,Ben-Dayan:2014lca,Ruehle:2015afa,Gao:2014uha} for example and see \cite{Baumann:2014nda} for a review) which would generically imply that the lightest field is a mixing of the two largest moduli, only one of which needs be a large modulus. 

Within a more general context we are therefore motivated to study properties of the Kahler potential of CY moduli for any two moduli whose axion partners may mix. Let us consider an example to illustrate the important physics mechanisms which come into play. Consider a Kahler potential of the type
\be
{\cal V}' = u_1^{\frac32} - u_2^{\frac32} - u_3^{\frac32} \;.
\ee
This has one large modulus and two small moduli. Say we now wanted to mix the $\nu_1$ and $\nu_2$ axions. Then we have
\be
\frac{K_1}{K_2} = -\left(\frac{u_1}{u_2}\right)^{\frac12} = \frac{q^1}{q^2} \;.
\ee
Already from this we can see why an enhancement of the effective axion decay constant can not work. Since $u_1>u_2$ we must have $\left|q^1\right|>\left|q^2\right|$. This means that we can only expect to obtain an enhancement for the decay constant appearing in the instanton involving $u_1$. But this instanton is always subdominant to the instanton involving the modulus $u_2$ which has an un-enhanced decay constant. More generally consider mixing the two 'small' moduli which gives
\be
\frac{f_2}{f_3} = \left(\frac{u^3}{u^2}\right)^{\frac14} = \left(\frac{q^3}{q^2}\right)^{\frac12} \;,
\ee
up to corrections suppressed by the ratio of the small cycles to the big cycle. Mixing between the axions is suppressed by small to large cycle ratios. We now see an inverse scaling of the decay constants with the flux parameters than (\ref{fratios}) which means that we can obtain an enhancement of the axion decay constant. However, like in the small and large modulus mixing case above we also have, say for $q^2 \gg q^3$, that $u_2 \gg u_3$ which means that the instanton with the sub-Planckian decay constant dominates over the super-Planckian instanton.

We see therefore a new physics mechanism appearing to censure the enhancement of the axion decay constant, in these cases it is actually possible to enhance the decay constant parametrically but then the super-Planckian instanton is forced to be sub-dominant to the sub-Planckian instanton. Therefore over one super-Planckian period we have many large sub-Planckian modulations, the period is therefore predominately sub-Planckian.
The physical effect is therefore the same in spirit to the previous one: the flux affects the moduli in such a way that the sub-Planckian instanton is dominant. 
Within the context of the WGC the light particle is now not associated to the instanton responsible for the inflaton potential, but to the instanton associated to the much smaller period modulations. This is the example of the strong version of the WGC coming in and closing the `loophole' of the weak version of the weak gravity conjecture \cite{Rudelius:2015xta,Brown:2015iha,Hebecker:2015rya}. It is interesting that such a mechanism is seen to manifest in this case. 

Indeed it is possible to make a general statement regarding the cases where the moduli that mix appear in isolation in the volume expression, up to corrections coming from mixing effects which for such scenarios are expected to be small. In such cases we have
\be
\left|\frac{K_{\lambda}}{K_{\sigma}}\right| = \left|\frac{q^{\lambda}}{q^{\sigma}}\right| = \left(\frac{u_{\lambda}}{u_{\sigma}}\right)^{\frac12} \;. \label{kratios}
\ee
However we now see that if we try to induce a hierarchy by taking $q^{\lambda} \gg q^{\sigma}$ then we require that $u_{\sigma} > u_{\lambda}$ so that the enhanced decay constant instanton dominates, while this is incompatible with (\ref{kratios}).

%%%%%%%%%%%%%%%%%%%%%%%%%%%%%%%%%%%%%%%%%%%%%%%%%%%%%%%%%%%%%%%%%%%%%%%%%%%%%
\subsection{Two moduli mixing}
%%%%%%%%%%%%%%%%%%%%%%%%%%%%%%%%%%%%%%%%%%%%%%%%%%%%%%%%%%%%%%%%%%%%%%%%%%%%%

What about the cases where there are multiple terms involving a modulus and substantial mixing between axions? The simplest case is the setup with only two moduli. For such scenarios it is useful to first introduce some general formalism for obtaining the effective axion decay constant. First we should diagonalise the kinetic terms for the axions by introducing a matrix $D$ such that $D^T K'' D = I$, where $K''$ denotes the matrix of second derivatives of the Kahler potential. Then we have the canonically normalised axion fields vector $c = D^{-1} \nu$ and the superpotential combination which appears is $\hat{q}^T c$ where $\hat{q} = D^T q$. Now we need to consider the axion combination orthogonal to the massive one $\hat{q}^T c$, which we label $\psi$. In terms of this light axion we can write the $c_i$ that mix as 
\be
c = \frac{L \hat{q}}{\left|\hat{q}\right|} \psi\;,\;\; L = \left(\begin{matrix}
0 & -1 \\ 1 & 0
\end{matrix}\right) \;.
\ee
The light axion decay constants appearing in the two associated instantons are then given by 
\be
f^{-1}_{\mathrm{eff}} = \frac{D L D^T q}{\left(q^T D D^T q \right)^{\frac12}} \;.
\ee
We can simplify this further by using the F-terms for the moduli and we can use the fact that $K_{\lambda} K^{\lambda \sigma} K_{\sigma} = 3$ for further simplification. Also starting from the property of ${\cal V}'$ that it is homogeneous degree $\frac32$ in the moduli we can reach the identity $u = - D D^T K'$ with $K'$ the vector of first derivatives. Using these results the expression simplifies to 
\be
f^{-1}_{\mathrm{eff}} = -\frac{1}{\sqrt{3}}D L D^{-1} u \;. \label{finv}
\ee
Note that $D = O^T \mathrm{diag}\left(r_{\lambda}^{-\frac12}\right)$ with $O$ an orthogonal matrix and the $r_\lambda$ being the eigenvalues of $K''$. Then the matrix $D L D^{-1}$ has unit determinant. So the inverse effective axion decay constants are a unit determinant transformation of the moduli fields. The first direct result is therefore that that we can not have all the effective decay constants large. This is a realisation of the WGC which states that there must be some instantons which involve a sub-Planckian decay constant for any axion.

We can also write 
\be
f^{-1}_{\mathrm{eff}} = -\frac{1}{\sqrt{3\mathrm{det} K''}} L K' \;. \label{finv2}
\ee
Since $\mathrm{det} K''$ must be non-vanishing, and is expected to be less than one as it is the product of the diagonalised `fundamental' decay constants, we need to have one of the $K'$ small to get an enhancement of an effective decay constant. Another interesting property of (\ref{finv2}) is that the fluxes $q$ do not appear in it. They do appear implicitly since the equality holds on the F-terms locus, however there is no explicit dependence and so (\ref{finv2}) can be taken as a statement purely on the moduli space of string theory. Certainly if (\ref{finv2}) does not allow decay constants larger than one for any value of the moduli then no such possibility exists. If there is a point in moduli space where it does allow for large decay constants then one needs to go back and check that it is compatible with the F-term solutions.

Let us calculate (\ref{finv2}) for an interesting example case. We consider the example based on the projective space $P_{[1,1,2,2,6]}$ studied in \cite{Candelas:1993dm} which has the volume expression
\be
{\cal V}' = \sqrt{u_1}\left(u_2-\frac{2}{3}u_1\right) \;. \label{volexam}
\ee
In this example for all values of the moduli there is always significant moduli mixing. Also there is a point in moduli space where $K_1 \rightarrow 0$ which leads us to expect an enhancement of the $f^{-1}_{\mathrm{eff},2}$ decay constant. Plugging the expression (\ref{volexam}) into  (\ref{finv2}) yields
\be
f_{\mathrm{eff},1} = \sqrt{\frac32}\frac{1}{u_1} \;,\,\,f_{\mathrm{eff},2} = -\frac{\sqrt{6}}{\left( u_2 - 2 u_1\right)} \;. \label{k10exa}
\ee
This shows that $f_{\mathrm{eff},1}$ must be smaller than one while  $f_{\mathrm{eff},2}$ can be large in the point in moduli space $u_2 - 2 u_1 \sim K_1 \rightarrow 0$. Note that this point can indeed be reached by taking $q^2 \gg q^1$. However since $u_2 > u_1$ the instanton involving  $f_{\mathrm{eff},1}$ always dominates over that of  $f_{\mathrm{eff},2}$. Hence, as in the cases studied above, we find no possible enhancement of the effective axion decay constant.

%%%%%%%%%%%%%%%%%%%%%%%%%%%%%%%%%%%%%%%%%%%%%%%%%%%%%%%%%%%%%%%%%%%%%%%%%%%%%
\subsection{Three moduli mixing}
%%%%%%%%%%%%%%%%%%%%%%%%%%%%%%%%%%%%%%%%%%%%%%%%%%%%%%%%%%%%%%%%%%%%%%%%%%%%%

The next level of structure is the case with three moduli and axions which mix. There is a technical and a conceptual difference from the case of two moduli. The technical one is simply that computing the diagonalisation of the kinetic matrix and the effective decay constants is much more challenging and demands significantly more computing time. The conceptual difference is that we are interested in mixing two axions which constrained how their associated superfields appeared in the superpotential. For the case of an extra modulus there need not be a constraint such as (\ref{kahler1}). Indeed given an arbitrary superpotential the modulus and its axion partner can be completely fixed in an unrelated way to the two moduli/axions which mix. There are therefore two classes of three-modulus setups: The first is when the extra modulus is also fixed with the relation (\ref{kahler1}), this is the case if the superpotential is linear for example. Of the three axions, one of them would then have to be fixed by a leading non-perturbative term, leaving an effective theory of the two lightest axions. The second setup is the more general one where the additional axion and modulus are fixed independently of the other moduli and axions. Let us consider the more general setup first and we will make some comments about the more constrained cases later.

The more general situation is one where the additional axions and modulus are fixed independently of the other moduli and axions. At first it may appear that it is difficult to say anything in such a setup since we do not know the value of the additional modulus and the decay constants depend on it. However recall that when we analysed the two axion case in the expression (\ref{finv}) the fluxes dropped out and we could write this as an equation purely in terms of the moduli. The trading of fluxes for geometry can always be done by using (\ref{kahler1}). Once we have a purely geometric equation we can make a statement over the whole of the moduli space, independently of what superpotential fixes the moduli at those loci. This becomes a powerful tool when there are more than two moduli because we do not need to know how the other moduli get fixed, only that they lie in the interior of the moduli space. We will see that for some cases this is strong enough to rule out super-Planckian enhancements. However for other cases there are loci in moduli space which could lead to enhancements and whether they do or not depends on the form of the superpotential. We study examples of these cases in turn.

\begin{figure}[!htb]
\centering
\includegraphics[scale=.7]{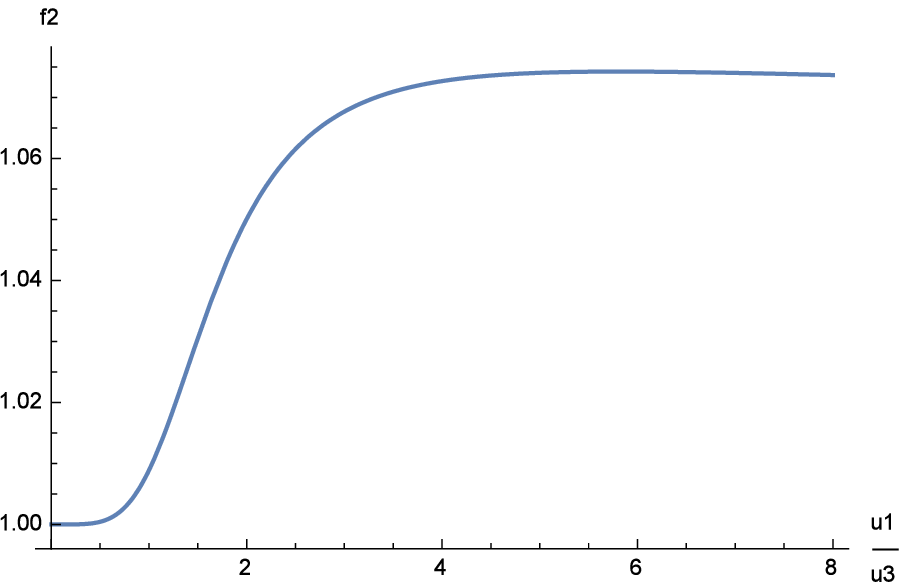}
\includegraphics[scale=.7]{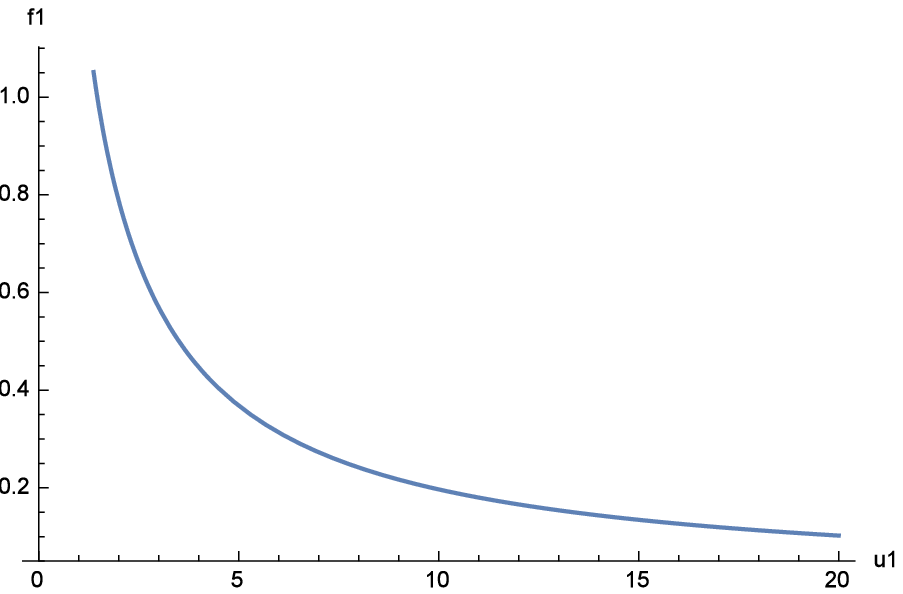}
\caption{Plots showing the effective axion decay constants for the two instantons associated to $u_1$ and $u_2$ as a function of the moduli. The $f_2$ plot is set at $u_2=1$ and the $f_1$ plot at $u_3=1$.}
\label{fig:f2case1}
\end{figure}

Let us consider an example which is simple enough to be solved completely over all the moduli space. Take the volume expression
\be
{\cal V}' = \frac12 \sqrt{u_1 u_2 \left(u_1 + 2 u_3\right)} \;.
\ee
This is the limit of an example in \cite{Cicoli:2011it} where the additional blow-up modulus is blown down. We are interested in mixing $\nu_1$ and $\nu_2$ in the superpotential while $\nu_3$ and $u_3$ are fixed arbitrarily (we can take $\nu_3=0$). Then solving for the effective axion decay constants we find
\be
f_{\mathrm{eff},1} = \frac{g_1\left( \frac{u_1}{u_3}\right)}{u_3} \;,\;\; f_{\mathrm{eff},2} = \frac{g_2\left( \frac{u_1}{u_3}\right)}{u_2}  \;, \label{f1f2case1}
\ee
where $g_1$ and $g_2$ are some known functions which are not strongly varying so that the overall moduli factors in (\ref{f1f2case1}) dominate them. In particular $f_{\mathrm{eff},1}$ is maximised for $u_3=1$ while $f_{\mathrm{eff},2}$ is maximised for $u_2=1$. The two effective axion decay constants, on their respective maximum locus in $u_3$ and $u_2$, are plotted in figure \ref{fig:f2case1}. We therefore see that they are never super-Planckian at any values of the moduli within the interior of the moduli space. This adds another example of sub-Planckian effective decay constants but more importantly it shows that it is possible to prove this without assuming anything about how the additional modulus and axion are fixed.
We can also consider mixing say $\nu_1$ and $\nu_3$ while fixing $\nu_2$ and $u_2$ arbitrarily. Or also a different expression for the volume of the same type, say ${\cal V}'=\sqrt{u_1\left(u_2^2 + 3 u_2 u_3 + u_3^2 \right)}$ from \cite{Cicoli:2011it}. The results are qualitatively the same as the example case above.

Let us consider now a qualitatively different example. Although the equation (\ref{finv2}) is valid only for two moduli it nonetheless leads one to naturally consider setups where some of the $K_{\lambda}$ vanish. A natural three-modulus extension of (\ref{volexam}) is to include also the dilaton modulus. This leads to an effective volume expression
\be
{\cal V}' = \sqrt{s u_1}\left(u_2-\frac{2}{3}u_1\right) \;. \label{volexamdil}
\ee
Consider a setup where we wish to fix $u_2$ and $\nu_2$ through some additional superpotential contribution while mixing $s$ and $u_1$. Solving for $K_s/K_1=q^s/q^1$ we get
\be
\frac{s}{u_1} = -\frac{q^1 \left(2 u_1 - 3 u_2\right)}{3q^s\left(-2 u_1 + u_2\right)} \;,\;\; f_{\mathrm{eff},s} = \sqrt{\frac23}\frac{\left(8 u_1^2 - 8 u_1 u_2 + 3 u_2^2\right)^\frac12}{s \left(-2 u_1 + u_2\right)} \;,\;\;
f_{\mathrm{eff},1} = \frac{\sqrt{6} (8 u_1^2 - 8 u_1 u_2 + 3 u_2^2)^\frac12}{u_1 \left(2 u_1 - 3 u_2\right)} \;.\nonumber
\ee
Here we have integrated out the massive axion combination in the superpotential (taking $q^2=0$), and also $\nu_2$. The remaining light axion combination has the decay constants $f_{\mathrm{eff},s}$ and $f_{\mathrm{eff},1}$ appearing in the associated instantons. This setup is somewhat similar to (\ref{k10exa}) in that we can obtain an enhanced axion decay constant $f_{\mathrm{eff},s}$ by going to the appropriate point in moduli space $u_2-2u_1 \rightarrow 0$. However the key difference is that by simultaneously also taking $q^s$ large we can choose the ratio $s/u_1$ so that the super-Planckian instanton dominates. The key question in realising such a scenario is what is the additional superpotential contribution that fixes $u_2$ and $\nu_2$ and further can it fix $u_2$ close to $2u_1$ without affecting the moduli values used in analysing the decay constants above? The spirit of this work is very much about accounting for how the moduli are stabilised in calculating the axion decay constants and instanton contributions and therefore this scenario lies outside the scope of the controlled examples used. Nonetheless it provides a possible direction to pursue in avoiding the super-Planckian censorship seen in all the cases analysed so far. 

What happens if we do not assume an additional superpotential contribution and fix the modulus $u_2$ instead through the single combination which appears in the superpotential $q^s S + q^1 U_1 + q^2 U_2$, for example as is the case in a type IIA on CY setting? Then it is immediate that we face difficulties in realising the super-Planckian scenario. Before even considering moduli stabilisation we see a fundamental obstruction: we would like $f_{\mathrm{eff},s}$ to appear in the dominant instanton, but the second heavy axion combination is now fixed by the most dominant instanton, and so $\sigma$ would be the axion to be fixed rather than $\nu_2$. Integrating out $\sigma$ first would then just lead to the two-modulus case studied previously. Similarly since the interesting locus in moduli space is $u_2=2u_1$ we see that the $\nu_2$ axion is lighter than the $\nu_1$ axion and so integrating it out first is not consistent. If we integrate out $\nu_1$ first then we get a mixing between $\sigma$ and $\nu_2$ but, unlike $K_1$, $K_s$ and $K_2$ can never vanish and so we do not expect a decay constant enhancement, indeed we find
\be
f_{\mathrm{eff},s} = \sqrt{\frac32}\frac{1}{s} \;,\;\;\; f_{\mathrm{eff},2} = \frac{\sqrt{54}}{3 u_2 - 2 u_1} \;,
\ee
which are always sub-Planckian.

To summarise: in the case with the number of axions being larger than two the qualitatively new effect is that where the additional moduli and axions are fixed is a superpotential dependent statement. However we have shown that for certain cases, even with large mixing, it is possible to rule out a super-Planckian enhancement over all of the moduli space. For other cases there are loci where enhancement is possible and the question is: can the moduli be fixed to these loci and what ratios of the moduli magnitudes are forced by this fixing? In the case where the superpotential depends only on one combination of all the axions, for example in a IIA on CY setting, the additional axion direction fixing is determined by the magnitude of the moduli appearing in the non-perturbative terms. We showed that for an interesting example this situation implies sub-Planckian effective decay constants only. However for a more general superpotential, for example as in \cite{Hebecker:2015rya} or IIA with metric fluxes, the possibility remains open that the moduli could be fixed with a dominant super-Planckian instanton, though there is no explicit realisation of this scenario as yet.\footnote{Note that the cost of introducing a more complicated superpotential is that it becomes non-generic and more difficult to leave axion flat directions in the first place. See for example \cite{Blumenhagen:2014nba,Hebecker:2014kva,Blumenhagen:2015kja} for studies in this direction. In particular in \cite{Blumenhagen:2014nba} it was shown that in IIB leaving an unfixed axion involving the partner of the dilaton, as required in realising the particular scenario above, is very restrictive on the fluxes and superpotential.} 

%%%%%%%%%%%%%%%%%%%%%%%%%%%%%%%%%%%%%%%%%%%%%%%%%%%%%%%%%%%%%%%%%%%%%%%%%%%%%
\section{Non-supersymmetric vacua}
\label{sec:nonsusy}
%%%%%%%%%%%%%%%%%%%%%%%%%%%%%%%%%%%%%%%%%%%%%%%%%%%%%%%%%%%%%%%%%%%%%%%%%%%%%

Perhaps the most straightforward way to avoid the conclusions so far is to fix the moduli in a non-supersymmetric vacuum. Since the analysis performed relied on solving the F-terms it is not clear that the conclusions, and in particular the scaling (\ref{fratios}), will still hold. Note that we only assumed that the F-terms for the geometric partners of the axions are satisfied, this does not mean that supersymmetry in the whole vacuum is preserved. Therefore to be clear, in this section we will consider non-supersymmetric stabilisation in the sense that the F-terms of the geometric partners of the axions themselves are not necessarily vanishing. Performing an analysis of non-supersymmetric geometric moduli stabilisation is much more complicated than the supersymmetric case. In particular the Kahler potential enters in a much more involved way. As an initial investigation let us restrict to a Kahler potential which is torodial in nature in the complex-structure moduli
\be
K = K^T - \log \left(S+\bar{S}\right) - \sum^3_{\lambda=1} \log \left(U_{\lambda}+\bar{U}_{\lambda}\right)  \;.
\ee
Note that this is also expected to capture the behaviour of 'large' moduli in Calabi-Yau compactifications. For example, in the $P_{[1,1,1,6,9]}$ example we see that for $u_1 \gg u_2$ we have an effective Kahler potential $K = -3\log \left(U_1+\bar{U}_1 \right)$. It is convenient to group the complex-structure moduli and dilaton together $U_I = \left\{S,U_{\lambda}\right\}$ which allows us to write the F-term scalar potential as
\bea
V_F &=& e^K \left[ K^{i\bar{j}}F_i \overline{F}_{\bar{j}} + \left| W \right|^2     \right. \nonumber \\
&+& \left. \sum_{I=0}^3 4 u_I\left(u_I \left|W_I\right|^2 - \mathrm{Re} \left(W_I \overline{W} \right) \right)   \right] \;. \label{V}
\eea
At this point we need to distinguish between fields that mix leading to the light axion and other moduli. Let us split the index $I$ into $\left\{\alpha,a\right\}$ where $\alpha$ runs over the axions that mix, and $a$ runs over the other axions. Then the superpotential must only be a function of $\Phi \equiv \phi+i\nu= \sum_{\alpha} q^{\alpha} U_{\alpha}$.
Proceeding, we can write the equations of motion in the vacuum $\partial_{u_{\alpha}}V = 0$ as
\be
a x_{\alpha}^2 + b x_{\alpha} + c = 0 \;, \label{quadratic}
\ee
where we defined $x_{\alpha} = q^{\alpha} u_{\alpha}$ (no sum over $\alpha$) and
\bea
a &=& 8 \left|W_{\Phi}\right|^2  \;, \nonumber \\
b &=& \partial_{\phi}\left( K^{i\bar{j}}F_i \overline{F}_{\bar{j}} - \left|W\right|^2\right) + 4 \sum_I u_I \left[u_I \partial_{\phi} \left|W_{I}\right|^2- \partial_{\phi} \mathrm{Re}\left(W_I\overline{W} \right) \right] \;, \nonumber \\
c &=& - V_F e^{-K}\;. \label{abc}
\eea
Note that (\ref{quadratic}) is not a quadratic equation in the $u_{\alpha}$ in general. However its coefficients are independent of the free index $\alpha$ which means that we can write
\be
\left(x_{\alpha}-x_{\beta} \right) \left(x_{\alpha}+x_{\beta}+\frac{b}{a} \right)=0 \;. \label{ratns}
\ee
This equation is the generalisation in the non-supersymmetric case of the supersymmetric version (\ref{fratios}), which in this notation is simply $x_{\alpha}=x_{\beta}$. We see that a similar structure emerges also in this case. First, one branch of the solutions, at $x_{\alpha}=x_{\beta}$, always exists independent of the values of $b$ and $a$ and this branch, which need not be supersymmetric, precisely reproduces (\ref{fratios}) and so any enhancement is canceled. If the $x_{\alpha}$ are both much smaller than $b/a$ then the full set of solutions are restricted to this branch. The second branch of solutions, denoted the split branch, is such that if the $x_{\alpha}$ are large with respect to $b/a$ we again find a relation like (\ref{fratios}) which implies no decay constant enhancement. 

The difficult region to analyse is when $x_{\alpha} \sim b/a$. Over the split branch the maximum splitting of the $x_{\alpha}$ is set at $b/a$ and so we should consider the properties of this quantity. The largest uncertainty comes from the Kahler moduli F-terms. From (\ref{V}) and (\ref{quadratic}) it is reasonable to expect that $K^{i\bar{j}}F_i \overline{F}_{\bar{j}} \rightarrow \infty$ sends $x_{\alpha} \gg \frac{b}{a}$ which leads us to the same regime of no possible enhancement. However to be able to make more precise statements we need to restrict to particular cases. Let us consider the type IIA superpotential, then the expression simplifies to 
\be
\frac{b}{a} = - \frac{1}{4} \left[ 2\mathrm{Im} W^T + \mathrm{Im} \left(K^{i\bar{j}}K_{\bar{j}} W_i\right)\right] \;. \label{baiia}
\ee
One immediate background which we can see is if we turn off the fluxes $f_0$, $\tilde{f}^i$ and $f_i$ to leave only the constant term $\tilde{f}_0$. In this case $b$ vanishes exactly and this is a non-supersymmetric no-scale configuration for the Kahler moduli. However for such a no-scale background we find no solutions satisfying also (\ref{quadratic}), which is in line with the absence of such vacua in the pure CY case as expected from the results of \cite{Camara:2005dc,Hertzberg:2007wc}.
An interesting property of (\ref{baiia}) is that it has no explicit dependence on the H-flux as well as the $F_6$ flux and the dilaton and complex-structure moduli. It is therefore possible that taking a scaling limit of large NS compared to RR fluxes leads to a region $x_{\alpha} \gg \frac{b}{a}$. However to show this explicitly one would need a full solution to the system. The only quantitative limit we could find is where $F_i=0$. Using these equations, as well as $\partial_{\nu} V=0$, the equation (\ref{quadratic}) simplifies to a set of coupled quadratic equations. The complete solution set can be obtained and we find that the possible ratios of the $x_{\alpha}$ are in the range $2/5 \le x_{\alpha}/x_{\beta} \le 1$. Hence in this case we can not induce any significant enhancement. 

In summary we find that solutions with $\left|x_{\alpha}\right| \sim \left|x_{\beta}\right|$, which mean that there is no possible enhancement of the effective axion decay constant, emerge naturally also in the non-supersymmetric case. However due to the complications of non-supersymmetric vacua we can only observe this behaviour for certain choices, or in certain limits or simplified settings. There is no doubt that the statements made for the non-supersymmetric case are inevitably significantly weaker than the supersymmetric case. But the results do hint that at least some of the conclusions reached regarding the absence of decay constant enhancement for the supersymmetric stabilisation case could also extend to the non-supersymmetric case.

%%%%%%%%%%%%%%%%%%%%%%%%%%%%%%%%%%%%%%%%%%%%%%%%%%%%%%%%%%%%%%%%%%%%%%%%%%%%%
\section{Summary}
\label{sec:summary}
%%%%%%%%%%%%%%%%%%%%%%%%%%%%%%%%%%%%%%%%%%%%%%%%%%%%%%%%%%%%%%%%%%%%%%%%%%%%%

In this work we studied whether it is possible to obtain an enhancement of an axion decay constant using axion alignment in string theory. The explicit string theory embedding we considered is a compactification of type IIA string theory on a CY with flux. The axion alignment was induced through the flux, following an idea presented in \cite{Hebecker:2015rya}, thereby implementing a variation of the KNP mechanism. We find that within this type IIA setting, if we solve for the moduli in a supersymmetric vacuum, accounting for the effect of the fluxes on the moduli implied that it is not possible to enhance the effective axion decay constant from mixing axions with respect to both the original axion decay constants. Since these latter must be sub-Planckian so must the new effective constant. The statement was dependent on particular properties of the Kahler potential which we argued are satisfied in the type IIA on CY settings for general CY manifolds if the flux coupling to the dilaton is not taken large. The physical mechanism at play was that the flux which induces the alignment also controls the moduli in such a way that the fundamental decay constants are modified so as to precisely cancel the potential enhancement from the mixing effect. If the dilaton flux is taken large then we showed for particular examples and classes of CY manifolds that one still does not obtain a decay constant enhancement but due to a different mechanism outlined below. 

Following this result in the IIA on CY setting, we performed a more general analysis of the effect of moduli stabilisation of the superpartners of the axions. In particular we required that, as in the string type IIA CY setting, the superpartners of the axions are fixed while leaving the inflaton axion a flat direction to be subsequently lifted through non-perturbative effects. We found that if the geometric moduli are fixed with vanishing F-terms then the statement of possible enhancement of the axion decay constant can be mapped to certain properties of the moduli space of CY manifolds. By studying examples, or restricting to certain classes of CYs, we found that either the same physical mechanism that appeared in the IIA setting described above manifests or a new physical mechanism arises which censures the possible enhancement of the effective axion decay constant. The new mechanism was that the locus in moduli space which allowed for an enhancement of the effective decay constant in one of the instantons associated to the two original axions, rendering it super-Planckian, was also the locus where the other instanton, which had a sub-Planckian decay constant, was dominant in magnitude over the super-Planckian one. We were able to show this behaviour for non-trivial, but still relatively simple, examples as well as for general CYs where the moduli partners of the axions which mix appear as single powers in the expression for the `volume'. In many of these cases the results did not rely on an assumption about how the moduli not involved in the mixing are fixed: the results hold over the full moduli space.

We did present examples of CY geometries involving three moduli (including the dilaton) where the question of the possibility of an axion decay constant enhancement depends on the form of the superpotential. For these cases we were unable to rule out a possible realisation of enhancement for general superpotentials thereby leaving a possible escape from the sub-Planckian behaviour found in all the explicit models. Potential realisations of such a scenario could be type IIA compactifications on non-CY spaces, or the type IIB setting in \cite{Hebecker:2015rya}. It would be very interesting to explore this possibility further. Of course we are also unable to constrain qualitatively different moduli stabilisation and axion alignment settings than the flux induced ones studied here. For example one could also consider more general potentials than we studied and in particular employ the use of D-terms to try and fix the moduli in a different way.\footnote{Note however that this is a delicate matter since typically adding D-branes to generate D-terms induces new open-string moduli which must be fixed also and by construction a D-term implies losing an axion direction in fields space so one must be careful to leave the inflaton axion massless.} Also there are a number of quite different mechanisms proposed for generating super-Planckian decay constants, see for example suggestions using a large number of axions \cite{Dimopoulos:2005ac,Bachlechner:2015qja,Bachlechner:2014gfa,Bachlechner:2014hsa} or axion kinetic mixing \cite{Shiu:2015uva,Shiu:2015xda}. We note that if a way to enhance the effective axion decay constant to super-Planckian values could actually be implemented in string theory then the type IIA setting we studied could form an interesting starting point for constructing models of inflation due to an exponential hierarchy between the masses of the moduli and the axions.\footnote{There is a no-go theorem against inflation in these vacua \cite{Hertzberg:2007wc} but this does not account for the non-perturbative effects and in fact can be more simply avoided through perturbative $\alpha'$ corrections \cite{Palti:2008mg}.}

One of the important assumptions of the analysis described above was that the moduli superpartners of the axions have vanishing F-terms. This does not mean that the whole background has to be supersymmetric. However it is still a strong constraint. We therefore studied a toy model, based on a torodial Kahler potential, where we dropped the F-term requirement. We found that the vacua split into two classes or branches of solutions. In the first we find the precise cancellation of axion decay constant enhancement seen in the supersymmetric case. In the second, so called split branch, we find that the deviation from this behaviour is limited in magnitude by $b/a$ in (\ref{abc}). Due to the complicated nature of non-supersymmetric vacua it is difficult to give general results on the magnitude of this deviation. The best we could do was identify certain limits where it is found to be small. Therefore, although the statements for the non-supersymmetric case are inevitably significantly weaker than the simpler supersymmetric one, we find similar structures of cancellation of the decay constant enhancement emerging at least for certain simplifying choices and limits.

For the example cases and classes of CYs that we studied, both in the explicit type IIA string theory setting and in the more general analysis, the results we find present new non-trivial quantitative support for the manifestation of the strong version of the Weak Gravity Conjecture in string theory. In the first physical mechanism we found the flux inducing the mixing led the cycle associated to the super-Planckian instanton to shrink bringing down a tower of light states from the higher wrapping branes which reduce the axion period. In the second one we found that within the interior of the moduli space the super-Planckian instanton had to be sub-dominant to a different sub-Planckian instanton. The sub-Planckian instanton therefore corresponds to the expect light states. This presents an explicit realisation of string theory enforcing the strong version of the WGC and closing the `loophole' considered in \cite{Rudelius:2015xta,Brown:2015iha,Hebecker:2015rya,Brown:2015lia,Heidenreich:2015wga,Junghans:2015hba}. 

Finally in appendix \ref{sec:axionmon} we presented a simple toy model of a flux induced axion monodromy model where the backreaction of the axion vev could be accounted for. In this model we find that as the axion traverses a period, the length of the period in field space decreases. The net effect is that the total field distance traveled is not linear in the number of periods traversed but only logarithmic. A similar effect was recently shown in a different setting in \cite{Blumenhagen:2015qda}. We hope to report on more work in this direction soon \cite{us}.

{\bf Acknowledgments}
I thank Arthur Hebecker, Timo Weigand and Lukas Witkowski for extremely useful extended discussions. The work of EP is supported by the Heidelberg Graduate School for Fundamental Physics.
%I wish a very happy birthday to the wonderful Raphael and thank him for extended discussions even though they were not particularly useful for this work. 

\appendix

%%%%%%%%%%%%%%%%%%%%%%%%%%%%%%%%%%%%%%%%%%%%%%%%%%%%%%%%%%%%%%%%%%%%%%%%%%%%%
\section{A toy axion monodromy model}
\label{sec:axionmon}
%%%%%%%%%%%%%%%%%%%%%%%%%%%%%%%%%%%%%%%%%%%%%%%%%%%%%%%%%%%%%%%%%%%%%%%%%%%%%

In this paper we have shown that the effect of the fluxes on the moduli can crucially modify the axion periods with respect to the naive field-theory result of keeping them fixed. In this appendix we present a very simplified toy model which aims to capture a similar effect in an axion-monodromy model. In spirit the model we consider is a flux-induced monodromy model following \cite{Silverstein:2008sg,Marchesano:2014mla,Hebecker:2014eua,Blumenhagen:2014gta}. We will find very similar results to \cite{Blumenhagen:2015qda} which perform a similar analysis in a non-geometric type IIB setting. The important effect we will find, which is also present in the model of \cite{Blumenhagen:2015qda}, is that in this setting the axion field range only grows logarithmically rather than linearly with the number of periods.  The particular model we consider is a simplified version of the IIA on CY model
\be
K = - \log s - 3\log u - 3\log t \;, \;\;\; W = \tilde{f}_0 - i h_0 S - i q U + \frac{1}{6} f_0 T^3 \;.
\ee
Now the massive axion combination is $\rho=h_0\sigma+q\nu$ which has had its moduli space decompactified through the flux. It is an axion with a flux induced monodromy. The idea is now to attempt to capture the back-reaction of the axion vev on the moduli and in turn on the axion decay constant itself. As is clear from the form of the superpotential, we can always trade a single period's worth of axion vev for a shift in a flux, in this case its $\tilde{f}_0$. This is expected to be a general property of flux induced monodromy. So let us consider instead analysing the structure of the potential at fixed zero vev for the massive axion, but with general flux $\tilde{f}_0$. The backreaction effect is then captured by the moduli dependence on $\tilde{f}_0$. We then hold the axion fixed, $\rho=0$, and minimise the scalar potential with respect to the other fields which appear. 

It is possible to find a physical solution to the equations $\partial_s V=\partial_u V=\partial_t V=\partial_b V=0$. It takes the form
\be
s=0.4 \frac{\tilde{f}_0}{h_0} \;,\;\; u=1.1 \frac{\tilde{f}_0}{q} \;,\;\; t=2.0 \left(\frac{\tilde{f}_0}{f_0}\right)^{\frac13} \;,\;\; b=0\;.
\ee
The numerical factors are approximate, but the important property is the scaling of $s$ and $u$ with $\tilde{f}_0$. Note that the scalar potential evaluated at this vacuum takes the form
\be
V = -0.6 \frac{h_0q^3 f_0}{\tilde{f}_0^3} \;,
\ee
which is AdS. This is consistent with the no-go theorem of \cite{Hertzberg:2007wc} against solutions to $\partial_s V=\partial_t V=0$ which only holds for positive cosmological constant.
Let us take for generality and to simplify notation
\be
s = \alpha \frac{\tilde{f}_0}{h_0}  \;,\;\; u = \beta \frac{\tilde{f}_0}{q} \;.
\ee
We can now see the qualitative effect directly: increasing $\tilde{f}_0$ decreases the size of the decay constants $f_{\sigma}$ and $f_{\nu}$. This means that as the axion traverses an increasing number of periods the period length decreases. More precisely there are two period lengths for the massive axion combination according to how it appears in the two instantons involving $s$ and $u$. There are
\be
f^{s}_{\rho} =\frac{h_0}{4\alpha^2\tilde{f}_0} \left(4 \alpha^2 +\frac{4 \beta^2}{3} \right)^{\frac12} \;,\;\;\; f^{u}_{\rho} =\frac{3q}{4\beta^2\tilde{f}_0}\left(4 \alpha^2 +\frac{4 \beta^2}{3} \right)^{\frac12} \;.
\ee
The dependence on the $h_0$ and $q$ fluxes in not important for the effect we would like to describe. The crucial thing is the inverse scaling with $\tilde{f}_0$. Under a periodic shift in the axions, which is along the massive axion direction
\be
\sigma \rightarrow \sigma + h_0 M \;,\;\;\nu \rightarrow \nu + q M \;,
\ee
we have that
\be
\tilde{f}_0 \rightarrow \tilde{f}_0 - \left(h_0^2 + q^2 \right) M \;.
\ee
For simplicity we can take the initial flux to vanish $\left.\tilde{f}_0\right|_{M=0}=0$. Therefore for $M$ periods traversed by the massive axion we have $\tilde{f}_0= -\left(h_0^2 + q^2 \right)M$.
This means that traversing $M$ periods does not lead to a distance in axion field space of $M \left.f_{\rho}\right|_{M=1}$ but rather to 
\be
\left.f_{\rho}\right|_{M=1}\sum_1^M \frac{1}{M} =\left.f_{\rho}\right|_{M=1}\left( \log M + \gamma + \epsilon_M \right) \;.
\ee
Here $\gamma$ is the Euler-Mascheroni constant, and $\epsilon_M \sim \frac{1}{2M}$. We therefore see that this particular toy model of capturing the backreaction of the axion we find only a logarithmic, rather than linear, enhancement of the axion field distance from monodromy. 

The effect that we showed is only in a simple toy model. Further it is most likely not a stable vacuum but only a turning point in the other moduli. Also the important property of the model is that the backreaction of the axion/flux energy density on the moduli was substantial while it maybe possible to tune this to be small, see for example \cite{Hebecker:2014kva,Blumenhagen:2015kja,Blumenhagen:2015qda}. Nonetheless it shows an interesting effect which, if generalises to more sophisticated and involved axion monodromy models, could have important implications. We hope to report on results in this direction soon \cite{us}.

\newpage
%\bibliography{papers}  
%\bibliographystyle{custom1}

\end{document}